\documentclass[letter,traditabstract]{aa}
\usepackage[dvips]{graphicx}
\usepackage{txfonts}
\def\G35{G35.20--0.74\,N}
\def\I{IRAS\,20126+4104}

\def\Msun{\mbox{$M_\odot$}}
\def\Lsun{\mbox{$L_\odot$}}
\def\AMM{NH$_3$}

\def\CO{$^{12}$CO}

\def\HCOpI{\mbox{H$^{13}$CO$^+$}}
\def\HCNI{\mbox{H$^{13}$CN}}

\def\METH{CH$_3$OH}
\def\MCN{\mbox{CH$_3$CN}}

\def\HM{H$_2$}
\def\HII{H{\sc ii}}

\def\kms{\mbox{km~s$^{-1}$}}

\def\mic{\mbox{$\mu$m}}

\def\Mstar{\mbox{$M_\ast$}}

\def\Vsys{\mbox{$V_{\rm sys}$}}

\begin{document}
\title{
A candidate circumbinary Keplerian disk in \G35:\\ A study with ALMA
}
\author{\'A.\ S\'anchez-Monge \inst{1} \and R.\ Cesaroni \inst{1} \and
M.~T.\ Beltr\'an \inst{1} \and M.~S.~N.\ Kumar \inst{2} \and T.\ Stanke \inst{3}
\and H.\ Zinnecker \inst{4} \and S.\ Etoka \inst{5,6} \and D.\ Galli \inst{1}
\and C.~A.\ Hummel \inst{3} \and L.\ Moscadelli \inst{1} \and
T.\ Preibisch \inst{7} \and T.\ Ratzka \inst{7} \and
F.~F.~S.\ van der Tak \inst{8,9} \and S.\ Vig \inst{10} \and
C.~M.\ Walmsley \inst{1,11} \and K.-S.\ Wang \inst{12}}
\institute{ INAF, Osservatorio Astrofisico di Arcetri, Largo E. Fermi 5, I-50125 Firenze, Italy 
            \and
            Centro de Astrof\'isica da Universidade do Porto, Rua das Estrelas, 4150-762 Porto, Portugal
            \and
            ESO, Karl-Schwarzschild-Strasse 2, D-85748 Garching bei M\"unchen, Germany
            \and
            SOFIA Science Center, NASA Ames Research Center, Mailstop 232-12, Moffett Field, CA 94035, USA
            \and
            Jodrell Bank Centre for Astrophysics, School of Physics and Astronomy, University of Manchester, Manchester M13 9PL, UK
            \and
	        Hamburger Sternwarte, Gojenbergsweg 112, 21029 Hamburg, Germany
            \and
            Universit\"ats-Sternwarte M\"unchen, Ludwig-Maximilians-Universit\"at, Scheinerstrasse 1, 81679 M\"unchen, Germany
            \and
            SRON Netherlands Institute for Space Research, P.O.\ Box 800, 9700 AV, Groningen, The Netherlands
            \and
            Kapteyn Astronomical Institute, University of Groningen, The Netherlands
            \and
            Department of Earth and Space Science, Indian Institute of Space Science and Technology, Thiruvananthapuram, India
            \and
	        Dublin Institute for Advanced Studies (DIAS), 31 Fitzwilliam Place, Dublin 2, Ireland
            \and
            Leiden Observatory, Leiden University, P.O.\ Box 9513, 2300 RA Leiden, The Netherlands
}
\offprints{\'Alvaro S\'anchez-Monge, \email{asanchez@arcetri.astro.it}}
\date{Received date; accepted date}

\titlerunning{An ALMA study of \G35}

\abstract{
We report on ALMA observations of continuum and molecular line emission with 0\farcs4 resolution towards the high-mass star forming region \G35. Two dense cores are detected in typical hot-core tracers, such as \MCN, which reveal velocity gradients. In one of these cores, the velocity field can be fitted with an almost edge-on Keplerian disk rotating about a central mass of $\sim$18~$M_\odot$. This finding is consistent with the results of a recent study of the CO first overtone bandhead emission at 2.3~\mic\ towards \G35. The disk radius and mass are $\ga$2500~au and $\sim$3~$M_\odot$. To reconcile the observed bolometric luminosity ($\sim$$3\times10^4$~\Lsun) with the estimated stellar mass of 18~\Msun, we propose that the latter is the total mass of a {\it binary} system.
}



\maketitle

\section{Introduction}\label{sint}

While different scenarios have been proposed to explain the formation of high-mass (i.e. OB-type) stars (monolithic collapse in a turbulence-dominated core -- Krumholz et al.~\cite{krumholz2009}; competitive accretion driven by a stellar cluster -- Bonnell \& Bate~\cite{bonnellbate2006}; Bondi-Hoyle accretion -- Keto~\cite{keto2007}; see also the review by Zinnecker \& Yorke~\cite{ziyo}), all of them predict the formation of circumstellar disks. It is thus surprising that only a handful of disk candidates have been observed in association with massive (proto)stars. As a matter of fact, despite many observational efforts, convincing evidence of disks has been found only around early B-type (proto)stars, while circumstellar disks around O-type stars remain elusive (Wang et al.~\cite{wang2012}; Cesaroni et al.~\cite{cesaroni2007} and references therein). Moreover, a detailed investigation of the disk properties, comparable to that performed in disks around low-mass stars (e.g. Dutrey et al.~\cite{dutrey2007}) is still missing due to the large distances of OB-type (proto)stars and the limited angular resolution at (sub)millimeter wavelengths. With the advent of the Atacama Large Millimeter Array (ALMA) the situation is bound to improve dramatically, as resolutions $\ll$1\arcsec\ will be easily obtained.

With this in mind, we performed ALMA Cycle 0 observations of two IR sources containing B-type (proto)stars. These were chosen on the basis of their luminosities (on the order of $10^4~L_\odot$), presence of bipolar nebulosities/outflows, detection of broad line wings in typical jet/outflow tracers (SiO), and strong emission in hot molecular core (HMC) tracers (such as methyl cyanide, \MCN). Here we present the most important results obtained for one of the two sources, \G35.

\begin{figure}
\centering
\resizebox{8.0cm}{!}{\includegraphics[angle=-90]{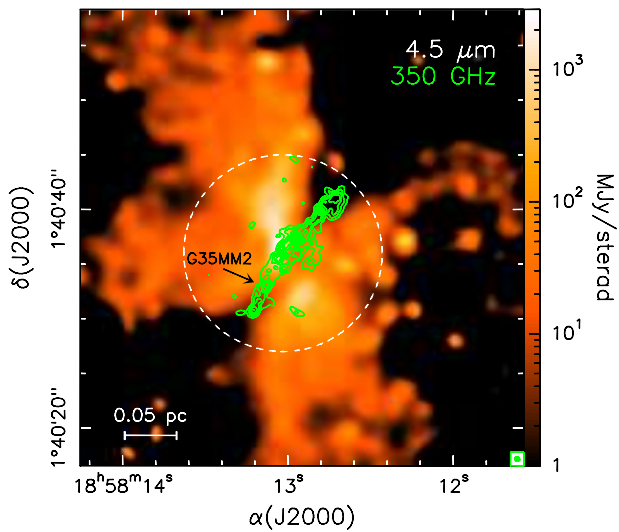}}
\caption{Large scale image of the 4.5~\mic\ emission towards the star forming region \G35\ overlaid with  a contour map of the 350~GHz continuum emission obtained with ALMA. The IR image has been obtained applying HiRes deconvolution (Velusamy et al.~\cite{velusamy2008}) to the Spitzer/IRAC data. The sub-mm map has been corrected for primary beam attenuation. The contours range from 9 to 203~mJy/beam in steps of 10.8~mJy/beam. The dashed circle denotes the primary beam of the ALMA 12-m antennas. The ellipse in the bottom right is the ALMA synthesized beam.}
\label{flarge}
\end{figure}

\G35\ is a well known star forming region located at a distance of $2.19_{-0.20}^{+0.24}$~kpc (Zhang et al.~\cite{zhang2009}), with a luminosity of $\sim$$3\times10^4~L_\odot$\footnote{Estimate taken from the RMS database, available at http://www.ast.leeds.ac.uk/cgi-bin/RMS/RMS\_DATABASE.cgi}.  The region is characterized by the presence of a butterfly shaped reflection nebula oriented NE--SW (see Fig.~\ref{flarge}), as well as a bipolar molecular outflow in the same direction, observed in \CO\ by Dent et al.~(\cite{dent1985a}), Gibb et al.~(\cite{gibb2003}; hereafter GHLW), Birks et al.~(\cite{birks2006}; hereafter BFG), and L\'opez-Sepulcre et al.~(\cite{lopezsepulcre2009}). The \CO(1--0) line emission appears to trace also a N--S collimated flow (see BFG), coinciding with a thermal radio jet (Heaton \& Little~\cite{heatonlittle1988}; GHLW) seen also at IR wavelengths (Dent et al.~\cite{dent1985b}; Walther et al.~\cite{walther1990}; Fuller et al.~\cite{fuller2001}; De~Buizer~\cite{debuizer2006}; Zhang et al.~\cite{zhang2013}). It has been proposed that the poorly collimated NE--SW outflow and the N--S jet could be manifestations of the same bipolar flow undergoing precession (Little et al.~\cite{little1998}). However, evidence for multiple outflows in this region is provided by SiO, \HCOpI, \HCNI, and \HM\ line observations (GHLW; Lee et al.~\cite{lee2012}).

A molecular clump elongated perpendicular to the NE--SW outflow has been mapped in dense gas tracers (\AMM, CS), whose emission exhibits a velocity gradient from NW to SE (Little et al.~\cite{little1985}; Brebner et al.~\cite{brebner1987}). This was first interpreted as a large ($\sim$1\arcmin\ or 0.6~pc) disk/toroid rotating about the NE--SW outflow axis, but GHLW, on the basis of their \HCOpI\ and \HCNI\ observations, propose that this is actually a fragmented rotating envelope containing multiple young stellar objects (YSOs). Indeed, GHLW identify a core at the center of the outflow and another core, named G35MM2, offset to the SE.

\section{Observations and results}\label{sobs}

\G35\ was observed with ALMA Cycle~0 at 350~GHz in May and June 2012, with baselines in the range 36--400~m, providing sensitivity to structures $\le$2~\arcsec. The digital correlator was configured in 4 spectral windows (with dual polarization) of 1875~MHz and 3840 channels each (covering the ranges 334.85--338.85~GHz and 346.85--350.85~GHz), providing a resolution of $\sim$0.4~km~s$^{-1}$.  Flux, gain, and bandpass calibrations were obtained through observations of Neptune and J1751$+$096. The data were calibrated and imaged using CASA.  A continuum map was obtained from line-free channels and subtracted from the data.  The synthesized beam is $0\farcs51\times0\farcs46$, P.A.=48\degr.  The rms noise is $\sim$6~mJy~beam$^{-1}$ for individual line channels, while in the continuum image it is $\sim$1.8~mJy~beam$^{-1}$, implying a S/N of only $\sim$100. The latter indicates a reduced dynamic range.

In Fig.~\ref{flarge}, we present the map of the 350~GHz continuum emission overlaid on an enhanced resolution Spitzer/IRAC image at 4.5~\mic\ extracted from the GLIMPSE survey (Benjamin et al.~\cite{benj}). The sub-millimeter continuum emission is clearly tracing an elongated structure across the waist of the butterfly shaped nebula. In all likelihood, we are detecting the densest part of the flattened molecular structure observed on a larger scale by Little et al.~(\cite{little1985}), Brebner et al.~(\cite{brebner1987}), and GHLW.

\begin{figure}
\centering
\resizebox{8.0cm}{!}{\includegraphics[angle=-90]{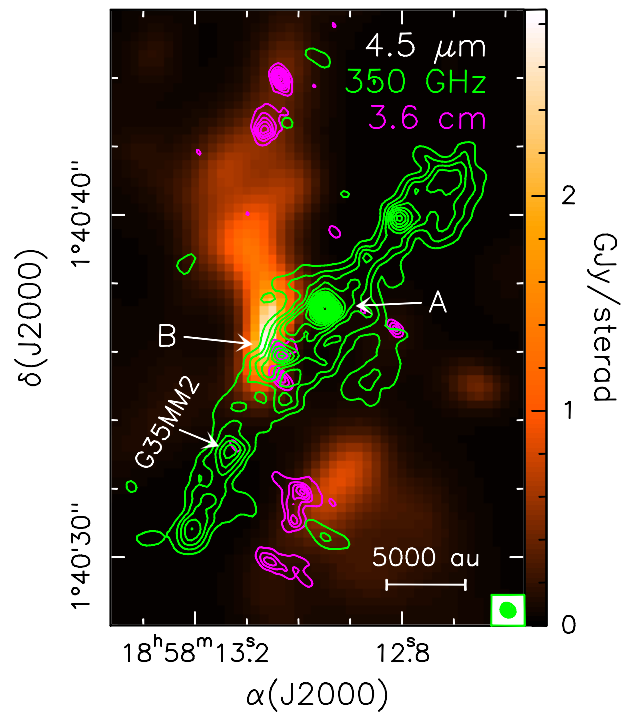}}
\caption{Same as Fig.~\ref{flarge}, with overlaid the 3.6~cm continuum map by GHLW (magenta contours). Unlike Fig.~\ref{flarge}, the 4.5~\mic\ image is displayed using linear scaling to emphasize the brightest structures outlining the N--S jet. Cores A, B, and G35MM2 are indicated by the arrows.}
\label{fcont}
\end{figure}

Along the elongated structure a chain of at least 5~cores is seen (Fig.~\ref{fcont}), lending support to GHLW's idea that one is dealing with a fragmented structure instead of the smooth disk/toroid hypothesized by Little et al.~(\cite{little1985}). We stress that the angular resolution of our maps ($\sim$7 times better than previous (sub)millimeter observations) reveals that the YSOs powering the outflow(s) lie inside cores A and/or B (see Fig.~\ref{fcont}), because these two are the only cores located close to the geometrical center of the bipolar nebula. In particular, core~B lies along the N--S jet traced by the IR and radio emission, and coincides with one of the free-free sources detected by GHLW. The hypothesis by BFG that core G35MM2 could be driving the NE--SW outflow is not convincing, as this core is off-center, lying right at the border of the waist outlined by the IR emission.

Methyl cyanide emission (as well as other typical HMC tracers) is clearly detected only towards cores A and B, and marginally towards G35MM2. Emission by vibrationally excited lines of \MCN\ also indicates that cores A and B could be hosting massive stars. Core B coincides also with a compact free-free continuum source detected by GHLW at 6 and 3.6~cm, and by Codella et al.~(\cite{codella2010}) at 1.3~cm. This emission could be part of the N--S thermal radio jet or might be coming from an \HII\ region ionized by an embedded early-type star. We will discuss this possibility in Sect.~\ref{sdis}. Faint radio emission at a $5.6\sigma$ level is detected also towards core~A, consistent with the presence of embedded star(s).

\begin{figure}
\centering
\resizebox{8.0cm}{!}{\includegraphics[angle=-90]{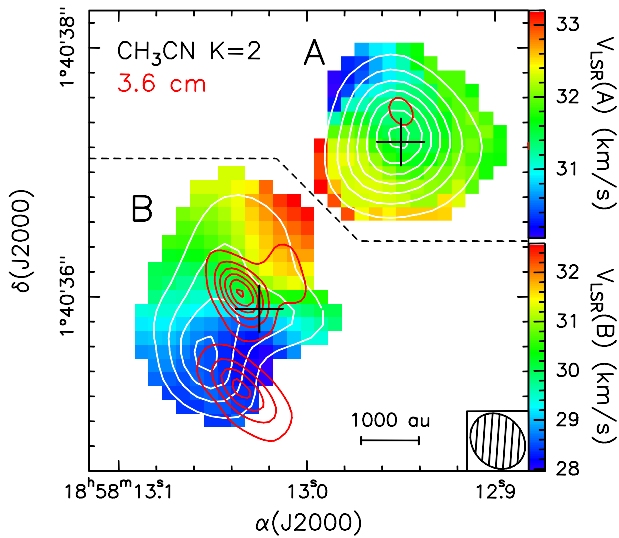}}
\caption{Overlay of the average emission in the \MCN(19--18) $K$=2 line (white contours) and 3.6~cm free-free continuum emission (red contours; from GHLW), on a map of the first moment (i.e. velocity) of the same \MCN\ line. The velocity intervals used to calculate the mean emission and first moment are 25.9--38.0~\kms\ for core A and 25.9--33.0~\kms\ for core B. The dashed line is intended to stress that different colour scales are used for the two cores to emphasize the velocity gradients. The white contour levels range from 80 to 656~mJy/beam in steps of 96~mJy/beam. The crosses mark the positions of the 350~GHz continuum peaks. The ellipse in the bottom right indicates the synthesized beam.}
\label{fvelo}
\end{figure}

We now investigate the gas velocity field in the two cores by computing the first moment of a \MCN\ line, a dense gas tracer. Figure~\ref{fvelo} plots the result for the \MCN(19--18) $K$=2 line, with overlaid the line emission averaged over the same velocity interval used to calculate the first moment. Note that the mean velocity of core~B ($\sim$30~\kms) differs by $\sim$2~\kms\ from that of core~A ($\sim$32~\kms), which also differs by a similar amount from that ($\sim$34~\kms) obtained with lower angular resolution observations of other tracers (see e.g. GHLW). Discrepancies of this type are often found in high-mass star-forming regions (see, e.g., the case of AFGL5142 -- Estalella et al.~\cite{estalella1993}; Zhang et al.~\cite{zhang2007}) and are likely related to the fragmentation process in molecular clumps. The interesting result is that both cores present velocity gradients, which may be due to expansion or rotation of the gas. Note that the sense of the gradient in core~B is the opposite of that measured on larger scales (see Fig~5 of Little et al.~\cite{little1985} and Figs.~8 and~9 of GHLW). However, the velocity range traced by the extended emission (30--41~\kms) differs significantly from that sampled by the \MCN\ lines ($\sim$26--38~\kms), suggesting that we are observing cores whose velocity field is not tightly related to that of the largest, fragmenting structure where they are embedded. In the following, we will address this question for core B, which is better resolved and associated with the N--S jet/outflow.

\section{The nature of core B: A Keplerian rotating disk?}\label{skep}

An elongated structure with a velocity gradient along its major axis, such as that observed in core~B, can be interpreted as either a collimated bipolar outflow, or a rotating ring/disk seen close to edge on. We thus first investigate the possibility that we are observing the root of a large-scale bipolar outflow.

At first glance, both the orientation (see Fig.~\ref{fcont}) and sign of the velocity gradient seen in Fig.~\ref{fvelo} appear consistent with those of the N--S jet imaged by BFG (see the red-shifted emission at 40~\kms\ in their Fig.~3). However, the CO emission in the northern jet lobe is detected by BFG also at 28~\kms\ (see their Fig.~3), namely at blue-shifted velocities. Moreover, the direction of the \MCN\ velocity gradient has a negative position angle, whereas the jet is slightly inclined to the east. Such a difference could be explained if the jet is precessing (as suggested by BFG) and lies close to the plane of the sky, because the jet direction would change from the small to the large scale.  We believe that this explanation is not satisfactory, because the orientation of the \MCN\ velocity gradient differs from that of the radio jet {\it on the same scale}. This can be seen in Fig.~\ref{fvelo}, where the two free-free emission sources are aligned N--S, whereas the velocity gradient (and the core major axis) are inclined to the west (PA$\simeq$--20\degr).

\begin{figure*}
\sidecaption
\includegraphics[width=14cm,angle=0]{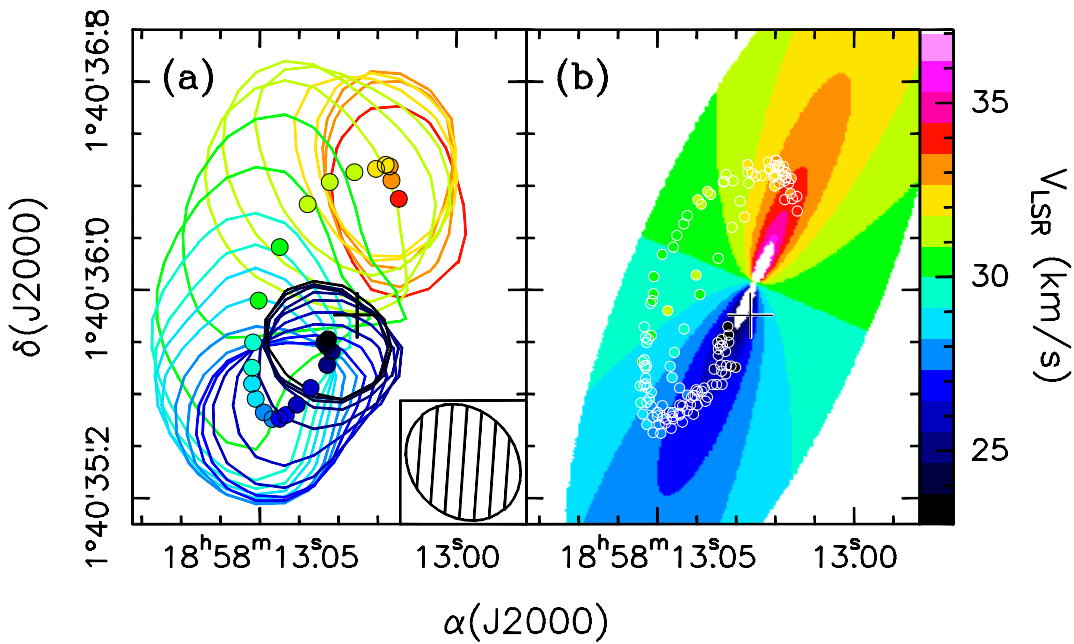}
\caption{{\bf(a)} Peaks of the \MCN(19--18) $K$=2 lines emission (solid circles) obtained with a 2-D Gaussian fit channel by channel. For each peak also the corresponding 50\% contour level is drawn (using the same colour as the peak). The colour corresponds to the line-of-sight velocity, according to the scale displayed to the right. The ellipse in the bottom right denotes the synthesized beam. {\bf(b)} Comparison between the velocity of the best-fit Keplerian disk (colour map) and the emission peaks at different velocities (solid circles) obtained for the following lines: \MCN(19--18) $K$=2,3,4, \METH($7_{1,6}$--$6_{1,5}$) $v_{\rm t}$=1, \METH($14_{1,13}$--$14_{0,14}$), HC$_3$N(37--36). The crosses mark the position of the continuum peak.}
\label{fkepfit}
\end{figure*}

Based on these findings, we consider the alternative possibility that the velocity gradient is tracing a disk. To analyse the kinematics of the gas in better detail, we fitted the \MCN\ emission in each 0.4~\kms\ velocity channel with a 2-D Gaussian. This allows us to derive the peaks of the line emission at different velocities. Then all peaks can be plotted together to obtain a picture of the gas velocity field. This is done in Fig.~\ref{fkepfit}a, where for each peak we also plot the corresponding 50\% contour level to give an idea of the size of the gas emitting at that velocity.

Two considerations are in order. The first is that the peaks outline a sort of elliptical pattern, roughly centered on the position of the continuum peak and oriented SSE--NNW. The second is that the most blue- and red-shifted peaks tend to converge towards the position of the sub-mm continuum peak. On this basis, one is tempted to hypothesize that the \MCN\ emission is tracing Keplerian-like rotation, as this would explain why both the emission at systemic velocities and that at high (blue- and red-shifted) velocities are observed towards the rotation center.

With this in mind, we have fitted the observed velocity pattern assuming Keplerian rotation about a point-like source. We stress that ours is a purely kinematical fit, where only the rotation velocity in a 2-D disk is calculated, and no estimate of the observed line intensity is computed. Also, no attempt is made to demonstrate the uniqueness of the model proposed and we envisage the possibility that other interpretations might be possible. Our simple model is sufficient to constrain a number of important parameters. The inputs of the model are: the LSR velocity of the system (\Vsys), the central point mass (\Mstar) and its position ($x_0$,$y_0$), the inclination angle ($\theta$) of the rotation axis with respect to the plane of the sky, and the position angle ($\psi$) of the projected major axis of the disk.
%

The best fit was obtained by varying all input parameters inside reasonable ranges and minimizing the expression
%
%
$\sum_i{\left({V_i^{\rm observed}}-V_i^{\rm model}\right)^2}$
where $i$ indicates the peak in a generic velocity channel. For our calculations, we have used the $K$=2, 3, and 4 components of the \MCN(19--18) transition as well as other prominent lines of other species, such as \METH($7_{1,6}$--$6_{1,5}$) $v_{\rm t}$=1, \METH($14_{1,13}$--$14_{0,14}$), and HC$_3$N(37--36). All of these lines reveal a kinematical pattern very similar to that in Fig.~\ref{fkepfit}a, demonstrating that such a pattern does not depend on the tracer and mirrors a real physical structure. In Fig.~\ref{fkepfit}b we show an overlay of the velocity peaks of all the lines and a velocity map of the best-fit model. Clearly, the agreement between the computed and observed LSR velocities is remarkable. The best-fit parameters are the following: \Vsys=30.0$\pm$0.3~\kms, $x_0$=18$^{\rm h}$58$^{\rm m}$13\fs027$\pm$0\fs002, $y_0$=01\degr40\arcmin35\farcs94$\pm$0\farcs07, $\psi$=157$\pm$4\degr, $\theta$=19$\pm$1\degr, \Mstar=18$\pm$3~\Msun.

A lower limit for the disk radius is $R_{\rm disk}\simeq2500$~AU, obtained from the maximum deprojected distance of the peaks from the center. One may wonder if such a big structure can undergo Keplerian rotation. The mass of the disk can be estimated from the continuum emission. The integrated flux density from core B at 350~GHz is 0.32~Jy. Assuming a dust absorption coefficient 0.5~cm$^2$\,g$^{-1}~(\nu/230.6~{\rm GHz})$ (Kramer et al.~\cite{kramer2003}) and a gas-to-dust mass ratio of 100, we obtain $\sim$3~\Msun\ for a dust temperature of 100~K. Despite the large uncertainties on the dust opacity and temperature, we believe that the mass of core B is significantly less than the central mass ($\sim$18~\Msun), which satisfies the condition for Keplerian rotation.

It is worth stressing that our findings are in good agreement with the recent study by Ilee et al.~(\cite{ilee}). Through measurements of scattered light from \G35, these authors find that the CO first overtone bandhead emission at 2.3~\mic\ can be fitted with a Keplerian disk rotating about a 17.7~\Msun\ star.

The distribution of the molecular peaks in Fig.~\ref{fkepfit}b clearly shows that our observations detect only the NE side of the disk. We speculate that this could be an opacity effect. If the disk is optically thick in the relevant lines, flared, and inclined by 19\degr, only part of the surface heated by the star is visible to the observer. This creates an asymmetry along the direction of the projected disk axis, with line emission being more prominent on the side (in our case the NE side) where the disk surface is visible.  Clearly, radiative transfer calculations are needed to confirm this scenario, but we note that in our source the NE part of the disk axis should be pointing towards the observer, consistent with the orientation of the CO outflow (blue shifted to the NE and red shifted to the SW - see GHLW) and the obscuration seen to the SW in the IR images (see Fig.~2).

\section{The stellar content of core B: A binary system?}\label{sdis}

An issue that is worth discussing is whether an 18~\Msun\ YSO is compatible with the bolometric luminosity of the region. ($3\times10^4~L_\odot$; see Sect.~\ref{sint}). Depending on the adopted mass--luminosity relation, the luminosity expected for an 18~\Msun\ star ranges from $2.5\times10^4~L_\odot$ (Diaz-Miller et al.~\cite{diazmiller1998}), to $6.6\times10^4~L_\odot$ (Martins et al.~\cite{martins2005}). This means that the 18~\Msun\ star should be the main contributor to the luminosity of the whole star forming region. Such a possibility seems quite unlikely due to the presence of multiple cores (see Fig.~\ref{fcont}), one of which is an HMC possibly hosting at least another high-mass star (core A).

A possibility is that one is underestimating the true luminosity due to the ``flashlight effect'', where part of the stellar photons are lost through the outflow cavities. According to the recent model by Zhang et al.~(\cite{zhang2013}), when this effect is taken into account, the luminosity obtained assuming isotropic emission ($3.3\times10^4~L_\odot$) becomes as large as $7\times10^4$--$2.2\times10^5~L_\odot$ consistent with a single star of $\sim$$20$--$34~M_\odot$.

While the previous explanation is possible, the isotropic estimate appears more robust than a model-dependent value, and we thus consider another hypothesis, namely that one is dealing with a binary system. In this case, the luminosity is significantly reduced with respect to that of a single 18~\Msun\ object and may be as low as $\sim7\times10^3~L_\odot$ for equal members. The latter is much less than the bolometric luminosity, thus allowing for a significant contribution by other YSOs.

The existence of a binary system could also explain why the N--S jet associated with core B is not aligned with the disk rotation axis. The presence of a companion would in fact be sufficient to induce precession of the jet/outflow, as hypothesized by Shepherd et al.~(\cite{shepherd2000}) to explain the observed precession of the bipolar outflow from the high-mass protostar \I. In this scenario, the outflow from core B would precess about an axis oriented NE--SW, i.e. along the bisector of the butterfly shaped IR nebula seen in Fig.~\ref{flarge}. The IR emission would arise from the cavity opened by the outflow itself during its precession, while the thermal radio jet would trace the current direction of the precessing axis.

The last question we address is the origin of the free-free emission from core B (see e.g. Fig.~\ref{fvelo}). Could this be tracing a (hypercompact) \HII\ region? According to GHLW, this source (n.~7 in their Table~1) has a spectral index $>$1.3 between 6 and 3.6~cm, compatible with free-free emission from an optically thick \HII\ region. Extrapolation of the 3.6~cm flux density (0.5~mJy) to 1.3~cm gives $>$1.9~mJy, in agreement with the 3~mJy flux measured, with lower angular resolution, by Codella et al.~(\cite{codella2010}). It is hence possible that the emission is partially thick at 1.3~cm and the Lyman continuum estimate of $10^{45}~{\rm s}^{-1}$ obtained under the optically thin assumption is a lower limit. A binary system with a total mass of 18~\Msun\ has a Lyman continuum flux (see Diaz-Miller et al.~\cite{diazmiller1998}) ranging from $5\times10^{44}$~s$^{-1}$ (for equal masses) to $1.8\times10^{47}$~s$^{-1}$. We conclude that we could be observing an \HII\ region ionized by a binary system at the center of a Keplerian disk. Should this be confirmed, \G35\ would represent a unique example of radio jet coexisting with an \HII\ region powered by the same YSO(s). Investigating such a short-lived transition phase in the evolution of an OB-type star could provide us with important clues on the formation process of these objects.

\begin{acknowledgements}
It is a pleasure to thank G\"oran Sandell for stimulating discussions on the \G35\ region and the anonymous referee for his/her constructive criticisms. We also acknowledge the support of the European ALMA Regional Center and the Italian ARC node. This paper makes use of the following ALMA data: ADS/JAO.ALMA\#2011.0.00275.S. ALMA is a partnership of ESO (representing its member states), NSF (USA) and NINS (Japan), together with NRC (Canada) and NSC and ASIAA (Taiwan), in cooperation with the Republic of Chile. The Joint ALMA Observatory is operated by ESO, AUI/NRAO and NAOJ. This work also used observations made with the Spitzer Space Telescope, which is operated by the Jet Propulsion Laboratory, California Institute of Technology under a contract with NASA. MSNK is supported by a Ci\^encia 2007 contract, funded by FCT (Portugal) and POPH/FSE (EC).
\end{acknowledgements}

\end{document}